\documentclass{article}
\usepackage{spconf,amsmath,graphicx,hyperref,booktabs,multirow}
\usepackage{enumitem}
\usepackage{placeins,float}
\setlist{nosep}

\title{The Universal Personalizer: Few-Shot Dysarthric Speech Recognition via Meta-Learning}
\name{Dhruuv Agarwal, Harry Zhang, Yang Yu and Quan Wang}
\address{Google DeepMind}

\begin{document}
\ninept
\maketitle
\begin{abstract}

Personalizing dysarthric ASR is hindered by demanding enrollment collection and per-user training. We propose a hybrid meta-training method for a single model, enabling zero-shot and few-shot on-the-fly personalization via in-context learning (ICL). On Euphonia, it achieves 13.9\% Word Error Rate (WER), surpassing speaker-independent baselines (17.5\%). On SAP Test-1, our 5.3\% WER outperforms the challenge-winning team (5.97\%). On Test-2, our 9.49\% trails only the winner (8.11\%) but without relying on techniques like offline model-merging or custom audio chunking. Curation yields a 40\% WER reduction using random same-speaker examples, validating active personalization. While static text curation fails to beat this baseline, oracle similarity reveals substantial headroom, highlighting dynamic acoustic retrieval as the next frontier. Data ablations confirm rapid low-resource speaker adaptation, establishing the model as a practical personalized solution.


\end{abstract}
\section{Introduction}
\label{sec:intro}

\subsection{\textbf{Motivation}}

Acoustic variability in dysarthric speech ~\cite{difficulty_dysarthria} creates a ‘utility gap’ that standard ASR systems~\cite{asr_survey} cannot bridge. Traditional personalization carries a prohibitive human cost: users with limited vocal stamina must often record hundreds of phrases to train dedicated models or adapters~\cite{tomanek2021residualadaptersparameterefficientasr}, creating a barrier for those with degenerative conditions who need the technology most. Aligning with the Interspeech 2026 theme of ‘Speaking Together,’ we propose a ‘Universal Personalizer’—a system that understands a unique voice in seconds, purely through context. By re-framing personalization as an inference-time task via In-Context Learning (ICL)~\cite{ZhuSunIclPersonalization}, we provide a few audio-text ‘shots’ to guide the model on-the-fly. This eliminates per-user training and gradient updates, enabling a truly universal model that provides immediate, high-quality ASR without the friction of data-heavy enrollment.

\subsection{\textbf{Related Work}}
Dysarthric ASR (DSR) research~\cite{qian2023survey} has explored fine-tuning and parameter-efficient methods like LoRA for individual speaker adaptation, with studies demonstrating personalization is possible even with small datasets (e.g., ~250 phrases per user in~\cite{tobin2022personalized}). Recent advancements include personalized RNN-T~\cite{graves2012sequence} approaches outperforming speaker-independent models by 35\%~\cite{Jimmy_single_asr_model_2024} on Euphonia~\cite{euphoniapaper}, and AdaLoRA~\cite{sap_personalization} improving over fine-tuning by 24\%~\cite{zhang2023adaloraadaptivebudgetallocation} on the Speech Accessibility Project dataset~\cite{sap_paper}.
Recently, the Interspeech 2025 SAP Challenge established new benchmarks. Top-performing systems achieved these results through strategies like offline weight-merging and custom audio segmentation. While these optimization techniques and adapters effectively push performance boundaries, large multi-modal models offer a distinct alternative paradigm: inference-time adaptation through ``MetaICL"~\cite{min2022MetaICLlearninglearncontext} or meta-learning.

Our approach uses meta-learning to use contextual examples for on-the-fly personalization~\cite{ZhuSunIclPersonalization} without weight updates. This work investigates its applicability and scalability for dysarthric ASR personalization across diverse datasets, aiming for state-of-the-art performance. It also explores if a single model performs well with and without few-shot examples, analyzes MetaICL's influence on learning dynamics, and examines the role of example curation~\cite{wang2024learningretrieveincontextexamples} for efficiency in ASR, similar to its effect on machine translation~\cite{agrawal2022context}.


\subsection{\textbf{Our Contributions}}

The core contributions discussed in the paper are majorly:
\begin{enumerate}[wide]
    \item \textbf{Hybrid Training Strategy:} We propose and validate a mixed training setup using both 0-shot (query only) and 10-shot (support + query) examples, excelling in all evaluation setups.
    \item \textbf{Effectiveness on Diverse Datasets:} We test this on the biggest dysarthric sets -- Euphonia and Speech Accessibility Project (SAP). 
    We set new non-personalized SOTA benchmarks and narrow the performance gap with per-user models.
    \item \textbf{Effective Example Curation:} Demonstrate how our trained model leverages same-speaker support examples to do on-the-fly personalization. We also explore different example curation methods via text-based embeddings, while establishing a theoretical upper bound on curation.
    \item \textbf{Data Ablation:} We analyze learning dynamics, specifically the rates of speaker and domain adaptation.
    
\end{enumerate}

\section{Methodology}
\label{sec:method}

\begin{figure*}
    \centering
    \includegraphics[width=0.95\textwidth]{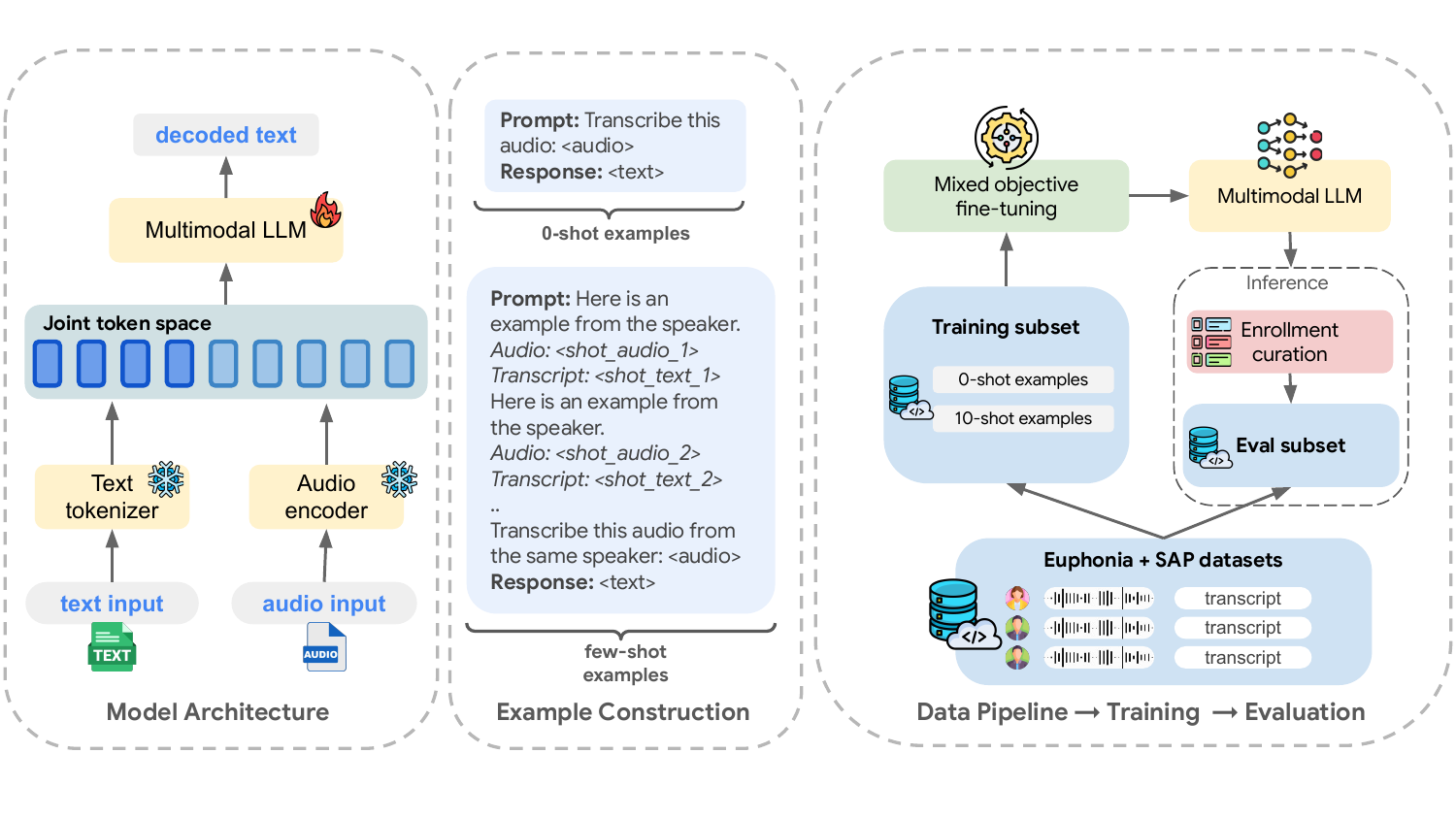}
    \caption{Overview. Left to Right: Architecture; Example construction; Mixed MetaICL training \& evaluation with example curation.}
    \label{fig:graphical_abstract}
\end{figure*}

\subsection{\textbf{Base Foundation Model: Gemini 2.5 Flash}}
Our work utilizes the Gemini 2.5 Flash model~\cite{comanici2025gemini25pushingfrontier}, a multi-modal LLM with advanced audio understanding and instruction-following capabilities. Its strong baseline instruction following performance aids the model to have strong capacity to learn with meta-learning fine-tuning. We only train the LLM layers while using the frozen audio tokenizer. The base model has not been trained on dysarthric speech. 
While we utilize Gemini 2.5 Flash for its incredible audio understanding capabilities, our MetaICL framework is model-agnostic and transferable to other multi-modal LLMs (e.g., Qwen-ASR\cite{shi2026qwen3asrtechnicalreport}, Llama-Omni 2~\cite{fangllama}) for similar on-the-fly personalization capabilities. However, our goal here was to showcase applicability and achieve best performance.


\subsection{\textbf{Meta-Learning for In-Context Personalization}}
We frame personalization as meta-learning, training a single model to recognize user-specific speech patterns via a small set of support examples provided in the prompt.
Our prompt consists of:
\begin{itemize}[wide]
    \item \textbf{Support Set}: $N$ pairs of support audio and corresponding text transcripts $\{ (a_{s_1}, t_{s_1}), ..., (a_{s_N}, t_{s_N}) \}$.
    \item \textbf{Query}: Single query audio utterance $a_q$ to be transcribed.
\end{itemize}
The model's task is to predict the transcript $t_q$ for query $a_q$, conditioned on the support set. The loss is calculated only on the query utterance. The training process optimizes the model's ability to perform in-context adaptation.

\subsection{\textbf{Datasets and Evaluation}}
We use two of the largest available dysarthric speech datasets.

\begin{itemize}[wide]
    \item \textbf{Google's Euphonia Project}: A large research initiative containing audio samples from more than 2,000 speakers with diverse etiologies (e.g., ALS, Parkinson's, vocal cord paralysis). We limited training to approximately 300k utterances (almost 30\% of the total). The test-set contains 5,684 utterances from over 199 speakers, reviewed by speech-language pathologists. For fair SOTA comparisons, we use a subset of 1,740 utterances.
    \item \textbf{Speech Accessibility Project (SAP)}: A collaboration led by the University of Illinois Urbana-Champaign and major tech companies to create a shared dataset for research. The training set contains 240k utterances, and the two distinct test sets (Test-1 and Test-2) have approximately 35k utterances each. 
    For SOTA evaluation, we utilize the 17k-utterance Test-1 subset established by the Interspeech 2025 Speech Accessibility Project (SAP) Challenge~\cite{zheng2025interspeech2025speechaccessibility}.
    Similar to Euphonia, we enforce strict speaker separation between splits.
\end{itemize}
\begin{table*}[th]
  \small
  \caption{WER (\%) comparison of different training strategies. We use complete test data available for Euphonia and SAP here.}
  \label{tab:main_results}
  \centering
  \setlength{\tabcolsep}{3pt}
  \begin{tabular}{l cccc ccc ccc}
    \toprule
    \multirow{2}{*}{\textbf{Model (Training Data)}} & \multicolumn{4}{c}{\textbf{Euphonia}} & \multicolumn{3}{c}{\textbf{SAP-Test-1}} & \multicolumn{3}{c}{\textbf{SAP-Test-2}} \\
    \cmidrule(lr){2-5} \cmidrule(lr){6-8} \cmidrule(lr){9-11}
    & \textbf{0-shot} & \textbf{5-shot} & \textbf{10-shot} & \textbf{19-shot} & \textbf{0-shot} & \textbf{5-shot} & \textbf{10-shot} & \textbf{0-shot} & \textbf{5-shot} & \textbf{10-shot} \\
    \midrule
    Baseline & 35.8 & 31.7 & 29.1 & 26.8 & 28.1 & 22.0 & 22.0 & 31.9 & 25.2 & 25.4 \\
    \midrule
    Euphonia 10-shot (20k) & 24.4 & 11.9 & 11.0 & 9.9 & - & - & - & - & - & - \\
    \midrule
    Euphonia 0-shot (200k) & 17.1 & 16.0 & 19.1 & 12.6 & - & - & - & - & - & - \\
    \midrule
    \begin{tabular}{@{}l@{}}Euphonia 10-shot, 0-shot (20k each)\end{tabular} & 17.1 & 13.6 & 11.1 & 9.8 & 17.8 & 14.3 & 13.7 & 20.0 & 16.1 & 15.3 \\
    \midrule
    \begin{tabular}{@{}l@{}}Euphonia \& SAP; 10-shot, 0-shot (20k each)\end{tabular} & \textbf{16.3} & \textbf{11.3} & \textbf{10.2} & \textbf{9.5} & \textbf{11.5} & \textbf{8.8} & \textbf{8.8} & \textbf{13.3} & \textbf{10.9} & \textbf{10.4} \\
    \bottomrule
  \end{tabular}
  \vspace{-1em} 
\end{table*}

\begin{table*}[htbp]
  \small
  \caption{Comparisons on published versions of test-sets, Euphonia and SAP, w.r.t speaker independent and dependent benchmarks.}
  \label{tab:sota_comparison}
  \centering
  \setlength{\tabcolsep}{3pt}
  \resizebox{\textwidth}{!}{
  \begin{tabular}{l l c c c}
    \toprule
    \textbf{Evaluation Set} & \textbf{Split} & \textbf{Our Model (N-shot ICL)} & \textbf{Non-personalized SOTA} & \textbf{Per-user SOTA}\\
    \midrule
    Euphonia (Severity) & (Mild, Moderate, Severe) & \textbf{(4.2, 11.2, 26.7)} (19-shot) & (7.3, 19.6, 31.3) (USM SI-ASR)~\cite{Jimmy_single_asr_model_2024} & -\\
    \midrule
    Euphonia (Subset) & Overall & 14.4 (10-shot);\textbf{13.9} (19-shot) & 17.5 (USM SI-ASR)~\cite{Jimmy_single_asr_model_2024} & \textbf{11.3} (RNN-T )~\cite{Jimmy_single_asr_model_2024} \\
    \midrule
    SAP Test-1 (Subset) & Overall & 7.5 (0-shot); \textbf{5.3} (10-shot) & 5.9 (Parakeet-tdt~\cite{zheng2025interspeech2025speechaccessibility}) 10.6 (FFT)~\cite{sap_personalization} & 8 (AdaLoRA)~\cite{sap_personalization} \\
    \midrule
    SAP Test-2 (Subset) & Overall & \textbf{9.5} (10-shot) & \textbf{8.1}(Parakeet-tdt~\cite{zheng2025interspeech2025speechaccessibility}) 10.0 (self-training)~\cite{wang2025selftrainingapproachwhisperenhance} & -\\
    \bottomrule
  \end{tabular}
  }
  \vspace{-1em} 
\end{table*}

 We ensure strictly disjoint speakers between training and test sets to prevent information leakage. Diversity of dysarthria etiology and severity distributions is discussed in (\cite{euphoniapaper}, \cite{tobin2025towards}, \cite{zheng2025interspeech2025speechaccessibility}). Performance is reported as Word Error Rate (WER \%).
\subsection{\textbf{Training Strategies}}
We investigate four training strategies:

\begin{enumerate}[wide]

\item \textbf{Baseline (SFT):} A general-purpose baseline model with no specific fine-tuning on dysarthric speech.
\item \textbf{0-Shot Trained (SFT):}
This represents standard LLM-based ASR training without explicit context utilization. We train on a corpus of 200k examples of single audio-text pairs, $(a_q, t_q)$.

\item \textbf{10-Shot Meta-Trained:}
Here we train on 20k ICL-formatted examples. Each example contains a query utterance and an additional support set of 10 audio-text pairs, all from the same speaker. This explicitly trains the model to utilize the provided context.

\item \textbf{Mixed-Objective (0+10 Shot) Trained:}
Our proposed solution, where we train the model on a mixture of 10-shot and 0-shot examples. We hypothesize that this model learns both to perform high-quality zero-shot transcription and to effectively utilize contextual examples.

\end{enumerate}
\subsection{\textbf{Optimization of In-Context Example Selection}}

Our default approach utilizes random utterances from the target speaker as support examples. Inspired by machine translation work~\cite{agrawal2022context,garcia2023unreasonableeffectivenessfewshotlearning}, we explore if curating ICL examples influences ICL performance. We investigate three categories of curation strategies to optimize this selection:

\begin{table}[t]
  \small
  \caption{Curation strategies comparison on Euphonia. Only ``Random Different Speaker`` doesn't use same-speaker examples.}
  \label{tab:curation_results}
  \centering
  \resizebox{\columnwidth}{!}{%
  \begin{tabular}{l c c}
    \toprule
    \textbf{Curation Strategies (5-shot)} & \textbf{Type} & \textbf{WER (\%)} \\
    \midrule
    Random (Different Speaker) & Baseline & 18.8 \\
    Random (Same-Speaker) & Our Default & 11.3 \\
    Diversity (Cluster Centroid) & Offline/Static & 11.0 \\
    Most Uncertain & Offline/Static & 11.4 \\
    Most Similar & Oracle (Dynamic indexing) & \textbf{9.9} \\
    \midrule
    19-shot Random (same speaker) & Upper Baseline & 9.5 \\
    \bottomrule
  \end{tabular}
  }
  \vspace{-1em} 
\end{table}

\begin{enumerate}[wide]
    \item \textbf{Random}: 
    We compare support sets comprised of (a) Random (different speaker) utterances  against (b) Random (same speaker) utterances; to isolate the impact of personalization on speaker-specific acoustics.
    \item \textbf{Computed Static Representative Sets}: We propose "offline" selection methods that pre-calculate a single, optimal set of $N$ examples per user to serve as a static representation of their speech. We do this using text-based methods on a set of utterances per user (enrollment set). First, we obtain sentence embeddings of transcripts, using the publicly available Universal Sentence Encoder model~\cite{universal_sentence_encoder_paper}.
    \begin{enumerate}
        \item \textbf{Diversity (Centroid)}: We cluster the user's sentence embeddings and select cluster centroids to maximize phonetic and lexical coverage within the $N$-shot support set.
        \item  \textbf{Sorted by Uncertainty}: 
        We perform a 0-shot pass on the user's enrollment data and select utterances with the highest WER. The rationale is that providing the model with its own "weak spots" (hard negatives) may better calibrate it to the user's specific dysarthric patterns.
    \end{enumerate}

    \item  \textbf{Oracle with text-similarity (Dynamic)}: While not deployable (as it requires the ground truth), this method establishes the potential theoretical lower bound of WER achievable through optimal example retrieval. For each query, we select support examples with the highest cosine similarity between their transcript embeddings and the query transcript. 
\end{enumerate}

\begin{table}[t]
  \small
  \caption{Data ablation: Euphonia WER (\%) vs increasing data}
  \label{tab:data_efficiency}
  \centering
  \begin{tabular}{c c c c c}
    \textbf{Training data} & \textbf{0-shot} & \textbf{5-shot} & \textbf{10-shot} & \textbf{19-shot} \\
    \midrule
    0.0\% & 35.8 & 31.7 & 29.1 & 26.8 \\
    2.0\% & 22.6 & 16.1 & 14.8 & 15.1 \\
    20.0\% & 19.9 & 12.8 & 11.5 & 10.7 \\
    40.0\% & 17.8 & 12.1 & 11.2 & 10.1 \\
    60.0\% & 17.5 & 13.5 & 11.0 & 10.2 \\
    100.0\% & 17.6 & 11.6 & 10.7 & 9.8 \\
    \bottomrule
  \end{tabular}
  \vspace{-1em} 
\end{table}


\section{Experiments and Results}
\label{sec:exp}

\subsection{\textbf{Training Strategy Comparison}}
Our primary experiments, detailed in Table \ref{tab:main_results}, evaluate the efficacy of different training strategies.

\subsubsection{Comparing Single-Objective Training on Euphonia}
When training exclusively on Euphonia, we observe a clear trade-off with single-objective models. The 0-shot model is trained on 200k examples with 1 utterance each, compared to 20k examples with 11 utterances each for 10-shot. 
The 0-shot model reduces baseline WER by 52\% but fails to leverage ICL examples, with performance degrading from 5-shot to 10-shot. Conversely, the 10-shot only model excels at personalization, reaching an impressive 9.9\% WER at 19 shots, but it's 0-shot WER of 24.4\% is significantly weaker. This highlights that single objective models are not optimal for both scenarios.

\subsubsection{Mixed-Objective and Multi-Dataset Synergy}
The next model, Euphonia based mixed-objective model, extends the 10-shot model with additional 20k examples for 0-shot . Successfully resolving the trade-off, it achieves a strong 0-shot WER of 17.1\%, nearly matching the 200k 0-shot only model, while having excellent and consistent few-shot performance (9.8\% at 19 shots). The model generalizes impressively, reducing the SAP-Test-1 0-shot WER by 36.6\% without training on it. Also, the consistent N-shot performance shows we don't need to train on varying count of support examples.

The advantages are further amplified when incorporating SAP data into the training mix in our final model, achieving the best results across all conditions. On Euphonia, the 0-shot WER improves to 16.3\% and the 19-shot WER to 9.5\%. Similar improvements on SAP, with the SAP-Test-1 0-shot WER plummeting to 11.5\% and the 10-shot WER to 8.8\%. This demonstrates that a mixed-objective, mixed-dataset approach yields the most robust and personalized DSR system.

\subsection{\textbf{State-of-the-Art Comparison}}
As shown in Table \ref{tab:sota_comparison}, our best model significantly advances the SOTA on standardized subsets. On Euphonia, we achieve dramatic WER reductions across all severity levels compared to the USM~\cite{USMpaper} SI-ASR speaker-independent baseline~\cite{Jimmy_single_asr_model_2024}, with WER for moderate dysarthria dropping from 19.6\% to just 11.2\%. While a fully personalized RNN-T model still achieves a lower WER on the Euphonia (11.3\%)~\cite{Jimmy_single_asr_model_2024}, our model (13.9\%) surpasses speaker-independent baselines (17.5\% WER).
On SAP Test-1, our model establishes a new state-of-the-art, achieving a 5.3\% WER (10-shot). This outperforms both previous personalized methods like AdaLoRA (8.0\%)~\cite{sap_personalization} and the Interspeech 2025 SAP Challenge winning system (5.97\%~\cite{zheng2025interspeech2025speechaccessibility}). On SAP Test-2, our model achieves 9.5\% WER. While this trails the absolute challenge winner (8.11\%~\cite{zheng2025interspeech2025speechaccessibility}), it comfortably surpasses the second-best submission (10.0\%~\cite{wang2025selftrainingapproachwhisperenhance}). Crucially, our approach achieves this top-tier performance on both sets using a single, unified architecture, completely avoiding techniques like complex model-merging and custom audio chunking utilized by the challenge winners.

\subsection{\textbf{Efficacy of Enrollment Curation}}

We conduct the experiments on the best model from Table \ref{tab:main_results} and discuss the results shown in Table \ref{tab:curation_results} in-depth below:

\subsubsection{\textbf{Validating On-the-Fly Personalization}}
The personalization mechanism is isolated by comparing random support utterances from \textit{same-speaker} versus \textit{different-speaker} sets.

Providing 5 mismatched support examples yields 18.8\% WER, a degradation compared to the 16.3\% 0-shot baseline from Table \ref{tab:main_results}. This ``interference effect'' confirms the model actively maps support acoustics to the query, failing when identities conflict.
Conversely, \textit{same-speaker} examples drop WER to 11.3\%. This relative 40\% drop validates that the model utilizes in-context audio for precise speaker adaptation rather than generic domain robustness.

\subsubsection{\textbf{Static Curation Strategies}}

We evaluated if "Computed Static Sets" (Diversity and Uncertainty) could outperform random selection (Table \ref{tab:curation_results}). We find that random same-speaker selection is a robust baseline (11.3\%) which static curation fails to surpass. This indicates the primary driver of personalization here is acoustic consistency rather than the semantic diversity of the support set.

\subsubsection{\textbf{The Efficiency Headroom}}
Finally, our oracle text-similarity experiment demonstrates the theoretical upper bound of our approach. Semantically similar 5-shot examples reduces WER to 9.9\%, close to the performance of a 19-shot random baseline (9.5\%) with only a quarter of the few-shot data. This result highlights a significant "efficiency headroom." It indicates that while text-based static curation methods have saturated, dynamic retrieval strategies hold promise. This motivates future work into audio-based similarity indexing, which could approximate this oracle performance in a deployable, text-free manner.

\subsection{\textbf{Data Ablation: Domain vs. Speaker adaptation comparison}}

We conducted a data ablation study (Table \ref{tab:data_efficiency}), with 50k instances (40k 10-shot and 10k 0-shot). The loss is calculated only on the query utterance. The 4:1 ratio prioritizes the acquisition of the in-context learning mechanism while preserving sufficient zero-shot supervision and being optimal in training steps needed overall.
We see a distinct decoupling between the speaker and domain adaptation.

\begin{itemize}
    \item Rapid Mechanism Acquisition (Speaker Adaptation): With just 2\% or 1000 instances (800 10-shot \& 200 0-shot), the gap between 0-shot (22.6\%) and 10-shot (14.8\%) is fully established. This suggests that the logic of attending to user-specific context—is highly data-efficient and established early in training.
    \item Domain Adaptation comes next: In contrast, the 0-shot baseline improves till 40\% of data at 17.8\% (or 16k 10-shot \& 4k 0-shot examples) and then plateaus. This indicates that domain adaptation needed more training rounds but less than 50\% of our initial data is needed to have a competitive model.
\end{itemize}
This decoupling highlights that near-SOTA personalization capabilities can be achieved with a fraction of the data, provided the model has sufficient exposure to learn the ICL interaction pattern.

\subsection{\textbf{Inference Limitations vs. Operational Efficiency}}
We acknowledge that multi-modal LLM backbones require higher inference costs than lightweight RNN-T or LoRA-based models. However, user studies indicate that moderate-to-severe dysarthric speakers currently lack any functional ASR solutions. By achieving 11.2\% WER for moderate severity, we prioritize crossing this critical "usability threshold" over latency minimization, as the gain in fundamental utility outweighs the inference cost. Additionally, this approach optimizes operational efficiency by eliminating per-user training infrastructure, aligning with the foundation model paradigm: a single, universal model capable of instant, stateless personalization without the complexity of gradient-based updates.

\section{Conclusion}
\label{sec:conclusion}
This paper presents a robust and scalable solution for on-the-fly personalization of dysarthric speech recognition. 
Our key contribution is a mixed-objective meta-learning strategy that fine-tunes a multi-modal LLM to excel at both zero-shot transcription and few-shot, in-context personalization. 
We establish new state-of-the-art benchmarks on Euphonia and SAP Test-1, and achieve competitive performance on SAP Test-2, all without requiring user-specific weight updates, offline model-merging, or custom audio segmentation.


Our analysis of enrollment curation validates the model's active adaptation, evidenced by a 40\% WER reduction when using random same-speaker support examples. We find this random selection forms a surprisingly robust baseline that static, text-based curation strategies fail to beat. However, our oracle experiments reveal efficiency headroom when examples are matched via similarity, highlighting dynamic acoustic retrieval as the next critical frontier for this technology.


Lastly, our findings from the data ablation study show that the model learns to utilize in-context examples with only 1000 total instances, and domain adaptation although slower, also takes only 20,000 mixed objective instances to reach a desirable level.
This work opens the door for instant, high-quality ASR for the millions of individuals living with speech impairments, without the burden of a long and exhausting enrollment process.

\vfill\pagebreak

\bibliographystyle{IEEEbib}
\bibliography{strings}

@inproceedings{tobin2022personalized,
  title={Personalized automatic speech recognition trained on small disordered speech datasets},
  author={Tobin, Jimmy and Tomanek, Katrin},
  booktitle={ICASSP 2022-2022 IEEE International Conference on Acoustics, Speech and Signal Processing (ICASSP)},
  pages={6637--6641},
  year={2022},
  organization={IEEE}
}

@article{difficulty_dysarthria,
author = {Young, Victoria and Mihailidis, Alex},
year = {2010},
month = {06},
pages = {99-112; quiz 113},
title = {Difficulties in Automatic Speech Recognition of Dysarthric Speakers and Implications for Speech-Based Applications Used by the Elderly: A Literature Review},
volume = {22},
journal = {Assistive technology : the official journal of RESNA},
doi = {10.1080/10400435.2010.483646}
}

@inproceedings{zheng2025interspeech2025speechaccessibility,
  title={The Interspeech 2025 Speech Accessibility Project Challenge},
  author={Zheng, Xiuwen and Phukon, Bornali and Na, Jonghwan and Cutrell, Ed and Han, Kyu J and Hasegawa-Johnson, Mark and Jiang, Pan-Pan and Kuila, Aadhrik and Lea, Colin and MacDonald, Bob and others},
  booktitle={Proc. Interspeech 2025},
  pages={3269--3273},
  year={2025}
}

@article{qian2023survey,
  title={A survey of technologies for automatic Dysarthric speech recognition},
  author={Qian, Zhaopeng and Xiao, Kejing and Yu, Chongchong},
  journal={EURASIP Journal on Audio, Speech, and Music Processing},
  volume={2023},
  number={1},
  pages={48},
  year={2023},
  publisher={Springer}
}

@INPROCEEDINGS{Jimmy_single_asr_model_2024,
  author={Tobin, Jimmy and Tomanek, Katrin and Venugopalan, Subhashini},
  booktitle={ICASSP 2025 - 2025 IEEE International Conference on Acoustics, Speech and Signal Processing (ICASSP)}, 
  title={Towards a Single ASR Model That Generalizes to Disordered Speech}, 
  year={2025},
  volume={},
  number={},
  pages={1-5},
  doi={10.1109/ICASSP49660.2025.10888895}}

@inproceedings{euphoniapaper,
    title= {Disordered Speech Data Collection: Lessons Learned at 1 Million Utterances from Project Euphonia},
    author = {MacDonald, Bob and Jiang, Pan-Pan and Cattiau, Julie and Heywood, Rus and Cave, Richard and Seaver and others},
    booktitle={ISCA archive isca-archive.org},
    year	= {2021},
}

@article{sap_paper,
  title={Community-supported shared infrastructure in support of speech accessibility},
  author = {Hasegawa-Johnson, Mark and Zheng, Xiuwen and others},
  journal={Journal of Speech, Language, and Hearing Research. 2024 ; Vol. 67, No. 11. pp. 4162-4175},
  year={2024},
}

@inproceedings{sap_personalization,
  title     = {{Personalized Fine-Tuning with Controllable Synthetic Speech from LLM-Generated Transcripts for Dysarthric Speech Recognition}},
  author    = {Dominik Wagner and Ilja Baumann and Natalie Engert and Seanie Lee and Elmar Nöth and Korbinian Riedhammer and Tobias Bocklet},
  year      = {2025},
  booktitle = {{Interspeech 2025}},
  pages     = {3294--3298},
  doi       = {10.21437/Interspeech.2025-2155},
  issn      = {2958-1796},
}

@article{fangllama,
  title={LLaMA-Omni 2: LLM-based Real-time Spoken Chatbot with Autoregressive Streaming Speech Synthesis},
  author={Fang, Qingkai and Zhou, Yan and Guo, Shoutao and Zhang, Shaolei and Feng, Yang}
}

@inproceedings{wang2024learningretrieveincontextexamples,
    title = "Learning to Retrieve In-Context Examples for Large Language Models",
    author = "Wang, Liang  and
      Yang, Nan  and
      Wei, Furu",
    booktitle = "Proceedings of the 18th Conference of the European Chapter of the Association for Computational Linguistics (Volume 1: Long Papers)",
    month = mar,
    year = "2024",
    publisher = "Association for Computational Linguistics",
    url = "https://aclanthology.org/2024.eacl-long.105/",
    doi = "10.18653/v1/2024.eacl-long.105",
    pages = "1752--1767"
}

@inproceedings{ZhuSunIclPersonalization,
author = {Sun, Zhu and Feng, Kaidong and Yang, Jie and Qu, Xinghua and Fang, Hui and Ong, Yew-Soon and Liu, Wenyuan},
title = {Adaptive In-Context Learning with Large Language Models for Bundle Generation},
year = {2024},
isbn = {9798400704314},
publisher = {Association for Computing Machinery},
address = {New York, NY, USA},
url = {https://doi.org/10.1145/3626772.3657808},
doi = {10.1145/3626772.3657808},
booktitle = {Proceedings of the 47th International ACM SIGIR Conference on Research and Development in Information Retrieval},
pages = {966–976},
numpages = {11},
location = {Washington DC, USA},
series = {SIGIR '24}
}

@article{asr_survey,
  title={Automatic Speech Recognition: A survey of deep learning techniques and approaches},
  author={Ahlawat, Harsh and Aggarwal, Naveen and Gupta, Deepti},
  journal={International Journal of Cognitive Computing in Engineering},
  volume={6},
  pages={201-237},
  year={2025},
}

@article{graves2012sequence,
  title={Sequence transduction with recurrent neural networks},
  author={Graves, Alex},
  journal={arXiv preprint arXiv:1211.3711},
  year={2012}
}

@inproceedings{
   zhang2023adaloraadaptivebudgetallocation,
   title={Adaptive Budget Allocation for Parameter-Efficient Fine-Tuning },
   author={Qingru Zhang and Minshuo Chen and Alexander Bukharin and Pengcheng He and Yu Cheng and Weizhu Chen and Tuo Zhao},
   booktitle={The Eleventh International Conference on Learning Representations },
   year={2023},
   url={https://openreview.net/forum?id=lq62uWRJjiY}
}

@inproceedings{wang2025selftrainingapproachwhisperenhance,
      title={A Self-Training Approach for Whisper to Enhance Long Dysarthric Speech Recognition}, 
      author={Wang, Shiyao and Zhou, Jiaming and Zhao, Shiwan and Qin, Yong},
      year={2025},
  year      = {2025},
  booktitle = {{Interspeech 2025}},
  pages     = {3299-3303},
  doi       = {10.21437/Interspeech.2025-934},
}

@article{USMpaper,
  title={Google USM: Scaling Automatic Speech Recognition Beyond 100 Languages},
  author={Zhang, Yu and Han, Wei and Qin, James and Wang, Yongqiang and Bapna, Ankur and Chen, Zhehuai and others},
  journal={arXiv preprint arXiv:2303.01037},
  year={2023}
}

@inproceedings{garcia2023unreasonableeffectivenessfewshotlearning,
author = {Garcia, Xavier and Bansal, Yamini and Cherry, Colin and Foster, George and Krikun, Maxim and Johnson, Melvin and Firat, Orhan},
title = {The unreasonable effectiveness of few-shot learning for machine translation},
year = {2023},
publisher = {JMLR.org},
booktitle = {Proceedings of the 40th International Conference on Machine Learning},
articleno = {438},
numpages = {12},
location = {Honolulu, Hawaii, USA},
series = {ICML'23}
}

@inproceedings{agrawal2022context,
    title = "In-context Examples Selection for Machine Translation",
    author = "Agrawal, Sweta  and
      Zhou, Chunting  and
      Lewis, Mike  and
      Zettlemoyer, Luke  and
      Ghazvininejad, Marjan",
    booktitle = "Findings of the Association for Computational Linguistics: ACL 2023",
    month = jul,
    year = "2023",
    publisher = "Association for Computational Linguistics",
    url = "https://aclanthology.org/2023.findings-acl.564/",
    doi = "10.18653/v1/2023.findings-acl.564",
    pages = "8857--8873"
}

@article{comanici2025gemini25pushingfrontier,
      title={Gemini 2.5: Pushing the Frontier with Advanced Reasoning, Multimodality, Long Context, and Next Generation Agentic Capabilities}, 
      author={Comanici, Gheorghe and Bieber, Eric and Schaekermann, Mike and others},
      journal={arXiv preprint arXiv:2507.06261},
      year={2025},
}

@inproceedings{universal_sentence_encoder_paper,
    title = "Universal Sentence Encoder for {E}nglish",
    author = "Cer, Daniel  and
      Yang, Yinfei  and
      Kong, Sheng-yi  and
      Hua, Nan  and
      Limtiaco, Nicole  and
      St. John, Rhomni  and
      Constant, Noah  and
      Guajardo-Cespedes, Mario  and
      Yuan, Steve  and
      Tar, Chris  and
      Strope, Brian  and
      Kurzweil, Ray",
    booktitle = "Proceedings of the 2018 Conference on Empirical Methods in Natural Language Processing: System Demonstrations",
    month = nov,
    year = "2018",
    publisher = "Association for Computational Linguistics",
    url = "https://aclanthology.org/D18-2029/",
    doi = "10.18653/v1/D18-2029",
    pages = "169--174"
}

@inproceedings{tobin2025towards,
  title={Towards a Single {ASR} Model That Generalizes to Disordered Speech},
  author={Tobin, Jimmy and Tomanek, Katrin and Venugopalan, Subhashini},
  booktitle={ICASSP 2025-2025 IEEE International Conference on Acoustics, Speech and Signal Processing (ICASSP)},
  pages={1--5},
  year={2025},
  organization={IEEE}
}

@inproceedings{min2022metaicllearninglearncontext,
    title = "{M}eta{ICL}: Learning to Learn In Context",
    author = "Min, Sewon  and
      Lewis, Mike  and
      Zettlemoyer, Luke  and
      Hajishirzi, Hannaneh",
    booktitle = "Proceedings of the 2022 Conference of the North American Chapter of the Association for Computational Linguistics: Human Language Technologies",
    month = jul,
    year = "2022",
    address = "Seattle, United States",
    publisher = "Association for Computational Linguistics",
    url = "https://aclanthology.org/2022.naacl-main.201/",
    doi = "10.18653/v1/2022.naacl-main.201",
    pages = "2791--2809"
}

@inproceedings{tomanek2021residualadaptersparameterefficientasr,
    title = "Residual Adapters for Parameter-Efficient {ASR} Adaptation to Atypical and Accented Speech",
    author = "Tomanek, Katrin  and
      Zayats, Vicky  and
      Padfield, Dirk  and
      Vaillancourt, Kara  and
      Biadsy, Fadi",
    booktitle = "Proceedings of the 2021 Conference on Empirical Methods in Natural Language Processing",
    month = nov,
    year = "2021",
    address = "Online and Punta Cana, Dominican Republic",
    publisher = "Association for Computational Linguistics",
    url = "https://aclanthology.org/2021.emnlp-main.541/",
    doi = "10.18653/v1/2021.emnlp-main.541",
    pages = "6751--6760",
}

@misc{shi2026qwen3asrtechnicalreport,
      title={Qwen3-ASR Technical Report}, 
      author={Xian Shi and Xiong Wang and Zhifang Guo and Yongqi Wang and Pei Zhang and Xinyu Zhang and Zishan Guo and Hongkun Hao and Yu Xi and Baosong Yang and Jin Xu and Jingren Zhou and Junyang Lin},
      year={2026},
      eprint={2601.21337},
      archivePrefix={arXiv},
      primaryClass={cs.CL},
      url={https://arxiv.org/abs/2601.21337}, 
}

\end{document}